# Hierarchy of protein loop-lock structures: a new server for the decomposition of a protein structure into a set of closed loops

Simon Kogan[1*], Zakharia Frenkel[1, 2, **], Oleg Kupervasser[1, 3***] and Zeev Volkovich[2****]

[1]Genome Diversity Center, Institute of Evolution, University of Haifa, Haifa 31905, Israel
[2]Department of Software Engineering, ORT Braude College, Karmiel, Israel
[3]Transist Video LLC, Skolkovo 119296, Russia

*e-mail: simonkog@gmail.com
**e-mail: zakharf@research.haifa.ac.il
***e-mail: olegkup@yahoo.com
****e-mail: vlvolkov@braude.ac.il

**Abstract**

HoPLLS (Hierarchy of protein loop-lock structures) (http://leah.haifa.ac.il/~skogan/Apache/mydata1/main.html) is a web server that identifies closed loops - a structural basis for protein domain hierarchy. The server is based on the loop-and-lock theory for structural organisation of natural proteins. We describe this web server, the algorithms for the decomposition of a 3D protein into loops and the results of scientific investigations into a structural "alphabet" of loops and locks.

**Introduction**

Several years ago, it was revealed that the majority of known natural globular proteins can be considered to be combinations of closed loops of an approximately standard size (25-35 amino acids (aa)). This discovery was based on polypeptide chain statistics [1]. The discovery was applied toward the understanding of the protein folding mechanism [2, 3], protein structural organisation [2, 4] and protein evolution [5, 6]. It was proposed that closed loops of an optimal size were an important stage in protein evolution. At that particular stage, the natural proteins were small molecules (25-35 amino acids in size) in which the turning of the backbone back onto itself (i.e., the formation of closed loops) could be an important selective advantage that provided stability to the molecule. In the next stage, these small proteins made of one closed loop united to form modern-sized domains (50-300 aa). The main hypothesis is that the closed loops, being elementary protein modules, at least partially conserved their ancestral structural and functional properties in modern proteins. Presumably, such modules can be classified into a limited number of families, which each originate from a corresponding, early ancestor protein formed from a single closed-loop protein. Such types of primary conserved families have already been described [7, 8]. Thus, the presentation of proteins as a set of closed-loop, conserved, standard-sized modules would be very useful for protein characterisation and classification and for understanding protein evolution.

The first, pared-down version of the site was created in 2005. Recently, the site was considerably improved and a new, important, theoretical investigation related to the topic of the site was performed. The improvement and investigation were the reasons to write this paper.

The first paper about the web server (DHcL), which implements the decomposition of a protein into a set of closed loops, was published in 2008 [9]. This server demonstrates a set of the best loops (with the smallest distances between ends and allowing a small overlap of 5 aa) rather than the optimal decomposition of non-overlapping loops (as our web server does). We will discuss the differences in detail in the Results section.

In this paper, the manual for HoPLLS (Hierarchy of protein loop-lock structures) is given, and the applied algorithms are described.

We also present the results of the application of HoPLLS to the creation of a full library of protein-building, closed-loop elements. This data can be used to find a new structural alphabet [**7, 8**] based on the conserved modules and can be used for protein annotation and comparison.

**Methods**

  **1. Algorithms for decomposing the protein structure into loop-like elements**

In the loop-lock representation, we consider a protein to be a set of closed loops [**5-7**]. We name the place where the ends of a loop meet each other a van der Waals lock [**10, 11**]. The lock area is +/-Lr ("the lock radius" [**10**]) relative to the position of each end of the loop. Lr is equal to 1 amino acid, 2 amino acids or 3 amino acids.

In this paper, we present algorithms to decompose a protein into a set of *non-overlapping* loops. We implemented two loop decomposition approaches.

The first approach (the geometrical algorithm) is based on minimising lock distances (distance between two ends of the loop) and maximising coverage of the protein with loops. We give priority to loops with smaller distances between the Cα atoms at the ends of the loops and with longer lengths.

The second approach (the physical algorithm) is based on the loop density criteria, i.e., finding a set of loops with the maximum number of internal links (not only links at the ends). Internal links are pairs of Cα atoms separated by a small distance in 3D space and by a long distance along the protein. In both approaches, we bounded the loop size by minimal and maximal values.

Both the discussed algorithms are novel and were not considered previously.
We now describe these two algorithms in detail.

  **a. Geometrical criteria (geometrical algorithm)**

The procedure for the geometrically optimised solution is as follows:

1) First, we find the full set of potential loops (which possibly overlap with each other). The ends of a loop satisfy three conditions:

   (i) The distance between the ends ($Lw_j$, i.e., the distance between the Cα atoms at the ends of the loop) is shorter than a certain threshold ($Lw^{max}$);

   $$Lw_j < Lw^{max}, \quad (1)$$

   where $j$ is the number of the current loop;

   (ii) The distance between the ends is the local minimum of the distances within the area of the "lock radius";
   (iii) The sequence length ($Lp_j$) is limited by an upper threshold $Lp^{max}$ and a lower threshold $Lp^{min}$:

   $$Lp^{min} < Lp_j < Lp^{max},$$

(2)

where j is the number of the current loop.

2) For each selected fragment $j$, we define a weight:

$W_j = (1 - A) Lp_j / Lp^{max} + A (1 - Lw_j / Lw^{max})$,
(3)

where A defines the contribution of the lock distances relative to the sequence length of the loops.

The full, combined weight of a set of loops is the following:

$W(J) = \sum_{i=1}^{M} W_{Ji}$,
where
$J = [J_1, J_2, ..., J_M]$ is the set of the loop indexes.

3) We find a set $J_{best}$ of non-overlapping potential loops that have the maximum combined weight among all the possible non-overlapping loop sets.

$J_{best} = argmax_J (W(J))$,
(4)
*where argmax* is a function that gives the J value at which the function W(J) has its maximum.

This procedure is performed in linear time and follows the description in [**12**]. In fact, in this paper, we used a part of the algorithm described in chapter 2 ("A linear time maximum weight independent algorithm") of [**12**]. For some weighted interval graph, the algorithm searches for the maximum weight set of non-overlapping intervals with the maximum weight.

4) We find the second-best loop decomposition, the third-best loop decomposition, etc. (see below).

### b. Optimal loop density criteria (physical algorithm)

1) For a given three-dimensional protein structure, we find all the fragments (potential loops, which possibly overlap with each other) in which the distance between the ends is less than the threshold $Lw^{max}$; see Eq. (1) (The distance is not necessarily the local minimum in the distance space). The sequence length ($Lp_j$) is limited by the upper threshold $Lp^{max}$ and the lower threshold $Lp^{min}$; see Eq. (2).

2) We calculate the loop weight $WG_j$ which represents the number of amino acid pairs that are involved in the inter-loop interactions. Two amino acids are considered to be an inter-loop interacting pair if they fulfil the following conditions:

(i) Their Cα atoms are located closer than $Dw^{max}$;
(ii) The number of intermediate amino acids between the amino acids is larger than $Lm^{min}$;
(iii) Each amino acid in the loop can only participate in one pair;
(iv) We begin looking for the amino acid pairs from the first amino acid of the loop and move to the other end;

(v) The inter-loop interacting pair is the geometrically closest amino acids (except for the previously found amino acids) of the loop that satisfy the above conditions.

The full combined weight of a set of loops is the following:

$WG(J) = \sum_{i=1}^{M} WG_{Ji}$,
where
$J = [J_1, J_2, ..., J_M]$ is the set of the loop indexes.

3) We find a set $J_{best}$ of non-overlapping potential loops that have the maximum combined weight among all the possible non-overlapping loop sets.

$J_{best} = argmax_J (WG(J))$,
(5)
where *argmax* is a function that gives the J value at which the function W(J) has its maximum.

This procedure is done in linear time and follows the same above-mentioned description [**12**].

4) We find the second-best loop decomposition, the third-best loop decomposition, etc. (see below).

**2. Finding the sub-optimal loop sets: second-, third-, etc., best loop sets**

Let us consider a pool of all the possible loops for a given protein chain. The loops can be distinctive or can overlap with each other. We also have the optimal, i.e. maximal weight, set of non-overlapping loops from the loops belonging to the pool. Our objective is to find the second-best set of non-overlapping loops, the third-best set of non-overlapping loops and so on.

The second-best set must be different from the first set by at least one loop. Therefore, we can find the second-best set by excluding from the pool the loops that belong to the optimal set one by one; we then search, as described above, for the maximum-weight, non-overlapping loop set in the pool that is reduced by one loop each time. If there are N loops in the first (best) set, then we have N variants after excluding each loop once. The maximum weight set from the search over all the variants is the second-best loop set. We put the remaining sets that were not chosen into the collection of candidates. Note that if the second-best set is different from the optimal set by more than one loop (the *degenerate case*), we will obtain the same set a number of times equal to the number of different loops.

The third-best loop set should be different from the first and the second best sets by at least one loop. Hence we exclude from the pool the same loop that was excluded to obtain the second-best set. In addition, we exclude the loops belonging to the second best set, one by one, and renew the search for the maximum-weight, non-overlapping loop set. The third-best set is the set with the maximum weight from the results of the exclusions at this stage and from the sets within the collection of candidates. We add the sets produced in this stage that were not chosen to the collection of candidates.

If we have the above-defined degenerate case, the third-best set will be the same as the second-best set. In this case, we omit this result and begin to look for the next-best set.

The next-best sets that follow can be extracted by a similar procedure.

## 3. Recommended parameters values for the algorithms

We chose the parameters for the algorithms based on the optimal correspondence between an intuitive "human" decomposition and the decomposition with the algorithm. We recommend the following parameters with these criteria:

| Parameter | Recommended value | Physical meaning of parameter | Purpose |
|---|---|---|---|
| $Lw^{max}$ | 10 angstroms | It is the threshold distance between the ends of a potential loop. | It prevents a loop with a large distance between its ends. |
| $Lr$ | 3 amino acids | It is the lock area ("lock radius") relative to the position of each end of the loop. | It defines the exact positions of the loop ends. |
| $Lp^{min}$ | 15 amino acids | It is the lower threshold for the loop length. It is half the standard loop size, approximately 30 amino acids. | It prevents the appearance of loops that are too short. |
| $Lp^{max}$ | 45 amino acids | It is the upper threshold for the loop length. It is one and a half times the standard loop size, approximately 30 amino acids. | It prevents the appearance of loops that are too long. |
| $A$ | 0.8 | It defines the contribution to the weight function (geometrical method) of the lock distance relative to the sequence length of the loop; $0 \leq A \leq 1$ | A value close to 1 gives priority to loops with tight locks. A value close to 0 gives priority to longer loops. |
| $Dw^{max}$ | 10 angstroms | It is the maximal 3D distance between inter-loop interacting aa pairs. These pairs are used to calculate the weight function in the physical method. | Smaller values give priority to compact loops with many interacting pairs. |
| $Lm^{min}$ | 5 amino acids | Minimal number of aa along the protein sequence between the aa of an inter-loop | Prevents close neighbours in a sequence from being considered an |

|  |  |  | interacting pair. | interacting aa pair |
|---|---|---|---|---|
| $N_{best}$ |  | 5 | The number, chosen for the demonstration of the algorithm, of sets of best loops. | We demonstrate the five best sets of loops. |

## 4. HoPLLS implementation

Visualisation of HoPLLS results requires the Chime program to be downloaded (http://www.umass.edu/microbio/chime/getchime.htm). This program helps Internet Explorer visualise 3D molecular structures written in pdb format.

Let us explain the structure of our web server. Let us open the main page: http://leah.haifa.ac.il/~skogan/Apache/mydata1/main.html

The "menu" is located on the left side of the page. It includes:
I) **Main page** – gives links to the page for the loop structure calculation and to the Chime download site;
II) **Introduction** – explains the algorithm for finding the loop-lock structure;
III) **Theory** – contains the publications that are most relevant to the topic of the site: 3D protein structure and loop-lock decomposition;
IV) **Detect loop (geometrical algorithm)** – the main part of the site for the loop structure calculation;
V) **Example of output** – gives an example of the page that results from the "Detect loop" job for the protein 4tim.
VI) **References –** lists publications related to the topic of the site
VII) **Download C++ Program for finding loops** – our program for finding loops. We ask the reader to reference this paper if he uses this program for his task.
VIII) **Authors** – the creators of the site
IX) **Download Chime** – download Chime program
X) **Download Acrobat –** download Acrobat reader

Let us choose the most important option: the page of "**Detect loop (geometrical algorithm).**"

There are three ways to input a protein for the closed loop decomposition (Fig. 1):
I) to load the "pdb" file from the user's computer
II) to paste the pdb file in the "window" on the site
III) to input the PDB code (4 characters: 4tim, for example). The HoPLLS routine loads this protein from the Protein Data Bank (http://www.rcsb.org/pdb/home/home.do).

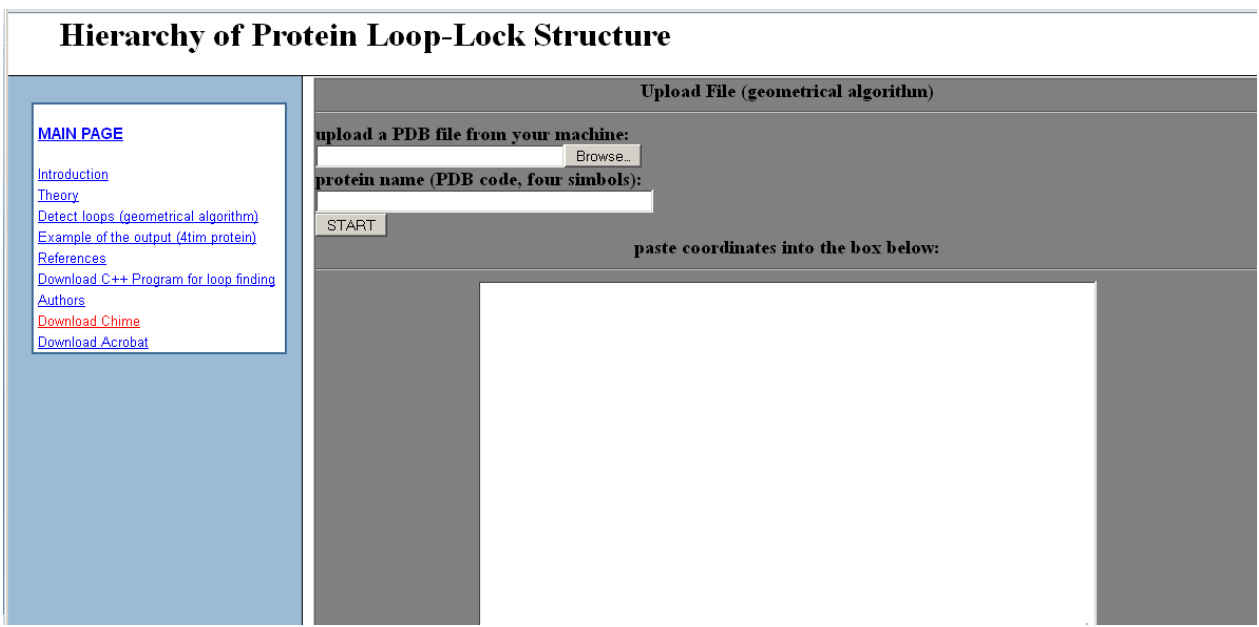

Fig.1 The "Detect loop" page to input a protein for detection of the loop-lock structure

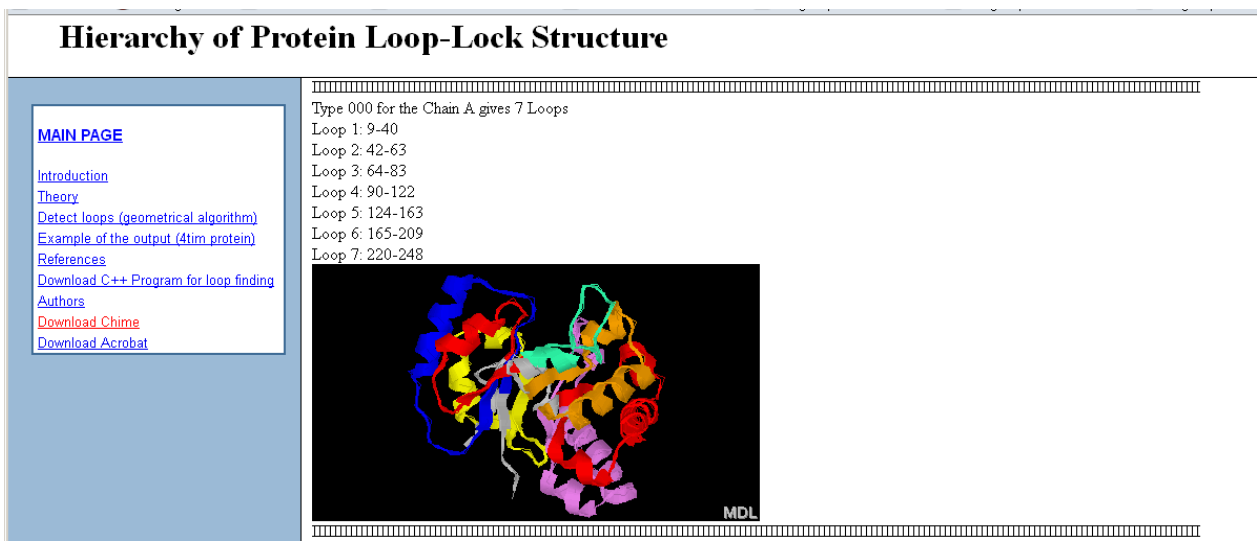

Fig. 2 Output page for the algorithm (for the loop-lock structural decomposition).

**Results**

### 1. Example of the output and comparison with DHcL

An **example of the output** page is shown in Fig. 2. The output includes the top five decompositions for every protein chain. The output lists the loops with the numbers of the first and the last amino acids; underneath, the image of the protein is given with the loops shown in different colours. The Chime program allows us to rotate the protein in 3D (in the image) with the computer mouse.

Consider for example the **4tim** protein (Chain A), which forms a TIM-barrel fold, decomposed by HoPLLS with the geometrical algorithm at the recommended parameters. The results for the closed-loop decomposition by the geometrical algorithm are shown in Fig. 3.

The results for the same task performed by the DHcL server [**9**] are shown in Fig. 4. This server uses a rather simple procedure for loop selection; it allows overlaps (five amino acids) and provides only one possible set of loops. This server repeatedly starts from the "better" loop (i.e., the tightest loop with the shortest Cα–Cα distance between the ends), and at each iteration, the sequence region corresponding to the mapped loop is excluded from further consideration.

In contrast, our server, HoPLLS, finds an optimal *non-overlapping* set by weighting the loops. In addition, HoPPLS provides several suboptimal sets through the algorithm for finding a Maximum Weight 2-Independent Set on Interval Graphs [**12**].

Although several of the selected loops are identical, some differences are also present. One of the main differences is the number of selected loops: 7 for HoPLLS and 8 for DHcL. This difference means that, indeed, the decomposition of the 3D protein structure into a closed loop assembly depends on the algorithm for loop selection. In the current case, the main reason for the difference is the following: we do not allow loops to overlap, whereas with DHcL, loop overlaps are possible. This difference explains the larger number of loops obtained with DHcL than obtained with HoPLLS.

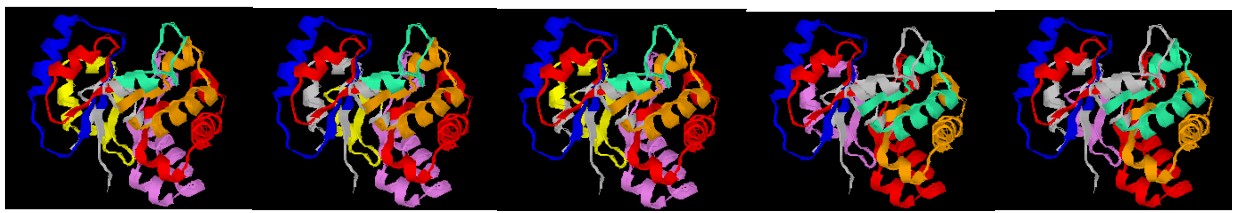

Fig. 3 Five sets for the loop decomposition of Chain A of **4tim**; the last two loop sets differ in Loop 6. The loop colours are: 1 blue, 2 red, 3 chlorine, 4 brown, 5 red, 6 violets, and 7 yellow.

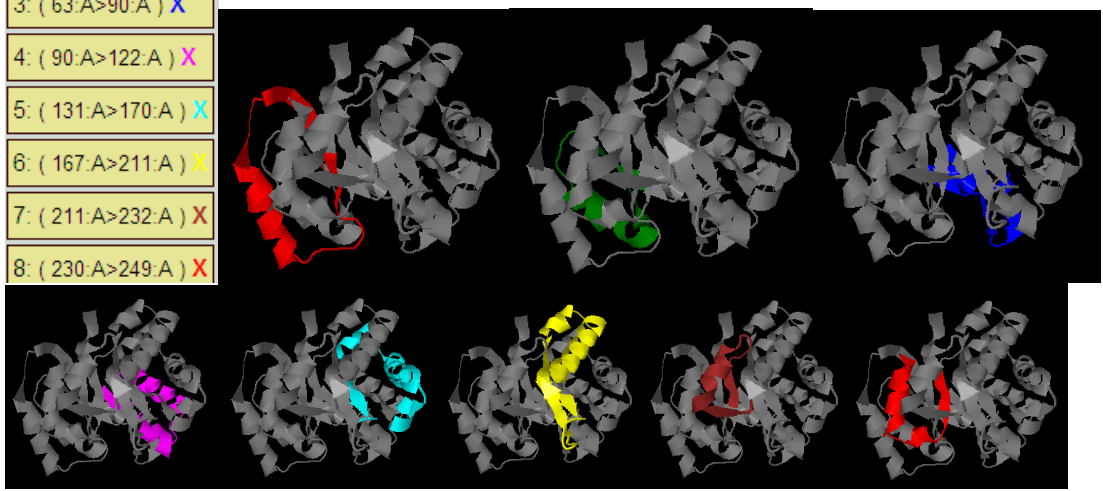

Fig. 4. The decomposition of the loops of Chain A from the DHcL [**9**] web server.

## 2. Closed-loop collections and looking for a universal protein code

Previously, it was proposed that a limited number of ancestral closed loops existed in the form of individual molecules and appear today as the structural elements of modern proteins, although their sequences and 3D structures have undergone significant changes during evolution. Detecting the different fragments in modern proteins that correspond to a closed loop ancestor is a challenging task.

There are generally two main approaches to detect this correspondence.

The first approach is from sequence to structure. Several studies have been published in this field [**7, 8, 13**], and several primary ancestral closed loops have been already described.

The second approach is from structure to sequence. This approach requires the extraction of loops from 3D protein structures prior to analysis of the sequences of these structures, as performed in [**14**].

We have applied our geometrical algorithm for loop extraction to approximately 6000 proteins from the PDB. These proteins have been decomposed into approximately 100,000 loops.

In our investigation, we planned to solve the following tasks:
(i) Form a library of all the possible loops from the proteins with known 3D structures.
(ii) Try to group most of the found loops into a small number of groups (with the same consensus sequence inside of every group), if such groups indeed exist. We name the set of the consensus sequences the "alphabet" of loops.
(iii) If a small number of such consensus sequences actually exist, we can find some correspondent sequence in a protein and predict the possible position of the loop from the 1D structure.

To select groups of similar loops, the following algorithm was used:

I) The sequence of the first loop in the whole set was compared to the sequences of all the other loops in the set. Two sequences were said to be close if they retained at least 30% similarity. The probability of this occurring randomly is very small and is equal to approximately 1/100,000 [**7**].
II) All loops that were sequence-wise similar to this loop were placed into the distinct group and were deleted from the set.
III) Return to the first step and repeat until the set is empty.

If such an "alphabet" of loops actually exists, they can be clustered into a small number of groups in such a way that all the loops in the same group are close to each other. Because the probability of random closeness is 1/100,000, in the absence of any "alphabet", we expect to form about 50,000 groups with a uniform distribution of the number of loops over the groups.

The number of detected groups was about 40,000. Most of the groups included two loops. The maximal number of loops in a group was fifteen. A more complex loop comparison model, which allowed insertions and deletions [**15**] of amino acids, was also applied. The number of groups found with this model was about 30,000. Most of the groups included three loops. The maximal number of loops in a group was twenty. We also used a "binary" code model [**16, 17**], in which 20 amino acids were divided into two groups that originated from the ALA and GLY amino acids. In the "binary" code case, two sequences were said to be close if they retained at least 90% similarity. Similarly to the 20 amino acids code, the probability of this occurring randomly in the "binary" code case is approximately 1/100,000. The number of groups found by the "binary" code was about 50,000. Most of the groups included two loops. The maximal number of loops in a group was nine.

The results demonstrate that the loop distribution is approximately random and that no "alphabet" of loops actually exists.

We also tried to find an "alphabet" of locks. The lock is defined to be the last three or five amino acids on both ends of the loop. The number of groups found in the lock-clustering experiments was found to be approximately the value expected for a random distribution: indeed, almost all combinations of amino acids were found with an approximately uniform distribution of lock numbers.

These results show that closed loop selection by our algorithm does not make it possible to decompose the complete set of loops into a small number of groups based on sequence similarity. This impossibility may exist because too many evolutionary changes have occurred in the primary closed loop ancestors. It is possible that more complicated procedures than direct sequence comparison should be applied.

It is well known that the same structure and even the same function often can be presented in nature by different sequences [**14, 18, 19**]; i.e., the sequence diversity in protein sequences is very high. A promising approach to overcome this problem is to define sequence relatedness in terms of connectivity through sequence similarity networks, as was proposed recently [**20**]. It was found that the natural sequence fragments form prolonged "walks" (connected through the networks), in which one sequence smoothly transforms to another one while conserving its structure and function. Remarkably, the sequences at the ends of such walks (on the periphery of the networks) are often completely different [**20, 21**]. Several examples of completely unmatched sequences from the same network are documented [**21, 22**], and many more can easily be found in the networks.

In many cases, these networks allow researchers to overcome the problem of sequence diversity and suggest a new criterion for sequence relatedness. In addition, these networks provide an alternative approach to all the methods in which notions of "sequence pattern" or "profile" and PSSM are used, e.g., [**23, 24**]. These notions imply an "average state" of sequences, whereas the network approach simultaneously considers a complete list, often of samples rather than different sequences linked to a common specific structure and function. This list may correspond to different "average states" yet be connected together through chains (walks) of pair-wise close similarities, which suggests evolutionary connections between the "average states". The same function for a given segment may be encoded by different consensus sequences. Moreover, in some cases a consensus may not exist at all.

**Conclusions**

In this paper, we described two algorithms for the closed-loop decomposition of 3D protein structures and a web server for the application of these algorithms. On the basis of the developed tools, a full collection of closed loops was investigated. The simple possibilities for finding an alphabet of loops and locks were verified. The future steps in this direction were considered.

**Acknowledgements**

We thank Prof. E.N. Trifonov for supervision during this project. This work has been supported by the European Union Seventh Framework Program through the PathoSys project (grant number 260429).